\documentclass[aps,twocolumn,groupedaddress,showpacs,floatfix]{revtex4}

\usepackage{graphics}
\usepackage{graphicx}
\usepackage{bm}
\usepackage{amsmath}
\usepackage{amsfonts} 
\usepackage{amssymb}
\usepackage{latexsym}
\usepackage{color}
\usepackage{mathrsfs}
\usepackage{braket}
\usepackage{enumerate}
\usepackage[usenames,dvipsnames]{xcolor}
\usepackage[colorlinks=true,citecolor=blue,linkcolor=blue,urlcolor=blue,backref=true]{hyperref}
\usepackage{float}

\begin{document}

\title{Putting Bell's inequalities into context by putting context into Bell's inequalities}
\author{Adam Stokes}
\affiliation{The School of Physics and Astronomy, University of Leeds, Leeds LS2 9JT, United Kingdom} 

\date{\today}

\begin{abstract}

Within the Dempster-Shafer theory of evidence a non-Kolmogorovian kind of epistemic uncertainty arises, which is encoded using multi-valued maps. We analyse the possible implications such non-Kolmogorovian epistemic uncertainty may have for Bell-type inequalities relating to the Einstein-Podolsky-Rosen-Bohm (EPRB) thought experiment. Our analysis leads to a notion of contextuality concerning complexes of physical measurement conditions. The use of multi-valued maps reveals an implicit link between this contextuality and counterfactual outcomes, and results in a formulation wherein the states of measurement devices are explicitly taken into account as part of the probabilistic event space. This reflects a conception of measurement that was advocated by Bell some time ago. It results in context-conditioned measure-theoretic probabilities, which do not obey Bell-type inequalities, but which are nonetheless perfectly compatible with local classical physical models. We give an example of a local classical model that reproduces the quantum mechanical predictions and that fits within the contextual framework.
\end{abstract}

\maketitle

\section{ Introduction}

Although we have recently celebrated the fiftieth year since Bell's original work \cite{bell_einstein-podolsky-rosen_1964}, Bell-type inequalities remain a contentious issue \cite{_journal_????}. From a mathematical perspective Bell-type inequalities merely express constraints on random variables defined over a single Kolmogorov probability space. It is quite remarkable that such seemingly innocuous and straightforward results of probability theory have produced such heated and protracted debates within physics \cite{pitowsky_george_1989,khrennikov_bell-boole_2008}. In the past Stapp claimed that Bell's inequality is {\em ``the most profound discovery of science"} \cite{stapp__1975}. On the other hand, more recently Khrennikov has claimed that {\em ``The only value of Bell's arguments was the great stimulation of experimental technologies for entangled photons."} \cite{khrennikov_contextual_2009}. It seems that those who agree with Stapp tend to be physicists, while those who agree with Khrennikov tend to be probability theorists.

In this paper we will argue that a {\em single} Kolmogorov space is very narrow as the basis for the treatment of realistic experiments. This limits the power of Bell-type inequalities in constraining the interpretation of physical models quite considerably. Our alternative approach utilises concepts from the Dempster-Shafer theory of evidence \cite{shafer_mathematical_1976,yager_classic_2008}. Unlike Kolmogorovian probabilities Dempster-Shafer probabilities are not additive, which makes the Dempster-Shafer theory more general \cite{shafer_mathematical_1976,yager_classic_2008}. The main use we find for this theory is in the identification of a non-Kolmogorovian kind of epistemic uncertainty. This uncertainty is associated with a probabilistic event space whenever two or more observables are operationally incompatible. It is encoded using multi-valued maps, which result in the replacement of a single Kolmogorov space with multiple Kolmogorov spaces, each labeled by distinct measurement contexts. Within the treatment of EPRB-type experiments, the multi-valued maps allow us to establish a link between measurement contexts and counterfactual outcomes. When hidden variables are considered the multi-valued maps result in a contextual approach similar to Khrennikov's \cite{khrennikov_contextual_2009,khrennikov_classical_2014}.

We begin in section \ref{eprb1} by considering Bell-type inequalities involving empirical data from a table. Such inequalities are based on phenomenological arguments alone and make no explicit reference to hidden-variables \cite{peres_unperformed_1978,peres_quantum_1995}. Instead they concern counterfactual outcomes of hypothetical measurements. Using the Dempster-Shafer theory enables us to probabilistically treat counterfactual outcomes differently to factual outcomes. The probabilistic uncertainty associated with the former is non-Kolmogorovian (despite being purely epistemic), while the uncertainty in the latter is of the usual Kolmogorovian variety. In section \ref{hv} we provide a treatment involving hidden-variables wherein multi-valued maps provide contextual observables and context-conditioned probabilities. These quantities are not constrained by Bell-type inequalities. In section \ref{c2} we relate our contextual approach to conventional approaches both rigorous and heuristic, and assess the significance of Bell's locality assumption. In section \ref{eg} we give an explicit example of a classical, deterministic, local model of an EPRB-type experiment that fits within our contextual approach, and that reproduces the quantum predictions. Finally, we summarise our findings in section \ref{conc}.

\section{The EPRB experiment}\label{eprb1}

In Bohm's version of the EPR experiment there are two spin-half particles produced by a common source. These particles travel in opposite directions labeled left and right. The spin of the left particle is measured using a Stern-Gerlach (SG) device aligned along one of two possible directions $a$ or $a'$, and the right particle is measured by a similar SG device aligned along either $b$ or $b'$. The generalised Bell-type CHSH inequalities refer to the following theorem \cite{clauser_proposed_1969}.

{\em Theorem}. Let $(\Omega,\Sigma,\mu)$ be a Kolmogorov probability space. Let real random variables $A, A':\Omega \to \{\pm\}$ represent the spin observables of the left particle along directions $a$ and $a'$ respectively, and let $B, B':\Omega\to \{\pm\}$ represent the spin observables for the right particle along $b$ and $b'$ respectively. The following Clauser-Horne-Shimony-Holt (CHSH) inequalities hold;
\begin{align}\label{chsh}
|f|:=|\langle AB \rangle + \langle AB'\rangle - \langle A'B\rangle +\langle A'B' \rangle| \leq 2
\end{align}
where the correlations are defined by
\begin{align}
\langle CC' \rangle := \int_\Omega d\mu(\omega) \, C(\omega)C'(\omega).
\end{align}
The proof is almost trivial, but is omitted for brevity.

\subsection{Phenomenological treatments and the Dempster-Shafer theory}\label{phen}

In a typical EPRB experiment measurements of the spins of an ensemble of $N$ particle pairs results in a table of values such as that given in \ref{t}. We let $\Omega_{\rm O}$ denote the set of $2N$ outcomes explicitly appearing in such a table. We denote by $\Omega_{\rm KPV}$ the set of values (KPVs) that are known to have been possessed by the particles at the time each particle was measured. Trivially, any reasonable theory of physics allows us to set $\Omega_{\rm O}\equiv \Omega_{\rm KPV}$.

A standard argument that a local deterministic classical theory cannot violate the CHSH inequalities proceeds as follows. Since, in a deterministic classical theory, the particles are assumed to possess spin values in all directions whether or not they have been measured, one can imagine that the blank entries in the table \ref{t} are in fact filled with variables which can take values $+$ or $-$. Let us denote by $\Omega_{\rm UPV}$ the set of {\em unknown} spin values possessed by the particles (UPVs). There are $2^{2N}$ concrete choices that can be made for $\Omega_{\rm UPV}$ as a set of $2N$ values each of which is either $+$ or $-$. These different choices are labeled $\Omega_{\rm CO}^i,~i=1,...,2^{2N}$. Each $\Omega_{\rm CO}^i$ represents a set of numbers called counterfactual outcomes.  If, contrary to the actual experiment that we are describing, we supposed that somehow all four spin directions $a,a',b,b'$ were measured for each particle pair, then we would obtain a complete table $\Omega^i = \Omega_{\rm O}\cup \Omega_{\rm CO}^i$ where the index $i$ takes a fixed value within $1,...,2^{2N}$. Each $\Omega^i$ comprises a complete table of outcomes, and each table can be viewed as a Kolmogorov space $\Omega^i$ in which the probability for an outcome is simply it's frequency of occurrence within the table.

\setlength{\tabcolsep}{10pt}
\begin{table}[t]
\begin{center}
\begin{tabular}{c|cc|cc}
       Run/ & \multicolumn{2}{c|}{Left particle} & \multicolumn{2}{c}{Right particle} \\
       Particle pair & $a$ & $a'$ & $b$ & $b'$ \\ \hline
       1 & $+$ & ? & $+$ & ? \\
       2 & ? & $+$ & ? & $-$ \\
       3 & $-$ & ? & $+$ & ? \\
       4 & ? & $-$ & $-$ & ? \\
       $\vdots$ & ~ & ~ & ~ & \\
       ~ & ~ & ~ & ~ & \\
       $\vdots$ & ~ & ~ & ~ & \\
       N & $-$ & $(+)$ & $(-)$ & $+$
\end{tabular}
\end{center}
\caption{A typical table of data collected in an EPRB experiment. Each run consists of one spin measurement on each particle. Each spin measurement is in one of two possible directions--- $a$ or $a'$ for the left particle, and $b$ or $b'$ for the right particle. The set of all entries is denoted $\Omega$. The set of all entries containing either $+$ or $-$ is denoted $\Omega_{\rm KPV}$. The set of all entries containing a ? is denoted $\Omega_{\rm UPV}$. In a complete table $\Omega_{\rm KPV}=\Omega$ and $\Omega_{\rm UPV}=\emptyset$, so in a complete table the unknown entries would also contain outcomes $\pm$. An example is given by the outcomes in brackets in the $N$'th row. Since such outcomes do not exist in any table of data produced by an actual experiment, they are termed counterfactual. There are $2^{2N}$ possible permutations of $2N$ variables $j=\pm$. Each permutation comprises a set $\Omega_{\rm CO}^i,~i=1,...,2^{2N}$. For fixed $i$ the set $\Omega_{\rm CO}^i$ gives one possible way to fill in the blank entries with counterfactual outcomes $\pm$.}\label{t}
\end{table}

However, attempting to encode local classical determinism by making a concrete choice for $\Omega_{\rm UPV}$ as one of the $\Omega^i_{\rm CO}$ conflates physical and probabilistic assumptions in way that may not be justified. Such an assumption is much stronger than classical determinism, because it assumes not only that there exist possessed values, but also that these values are known from having performed measurements. The assumption that particles can simultaneously possess definite spin values in different directions, does not by itself mean that we should assign relative frequencies to both KPVs {\em and UPVs}, as though {\em both} are measurement outcomes. Only KPVs should be treated as measurement outcomes. Put slightly differently, the assumption that possessed values merely exist does not mean that we should necessarily treat the two sets $\Omega_{\rm KPV}\equiv \Omega_{\rm O}$ and $\Omega_{\rm UPV}$ as part of the same probability space. Such a {\em probabilistic} restriction entails a {\em physical} constraint which is different to the one we seek to impose.

Within a single Kolmogorov probability space {\em all} probabilities whether they are taken to pertain to UPVs or to KPVs are required to satisfy the same rules. Since UPVs are unknown, their probability assignments must be {\em subjective}. On the other hand KPVs appearing explicitly in a table are afforded probability assignments in the form of {\em objective} relative frequencies. Clearly a single Kolmogorov space is too limited to even allow for the {\em possibility} that these two distinct types of probability might be treated differently. However, if we employ the Dempster-Shafer theory, such a distinction becomes possible. Subjective UPV-probabilities and objective KPV-frequencies are generally viewed differently. The uncertainty associated with UPVs is interpreted as uncertainty in the underlying event space and is non-Kolmogorovian, although it is entirely epistemic.

Rather than viewing table \ref{t} as defining $2^{2N}$ Kolmogorov spaces, we instead view it as defining a single Dempster-Shafer probability space $\Omega:=\Omega_{\rm KPV} \cup \Omega_{\rm UPV}$ in a sense that will become clear in what follows. The basic idea in our approach is to associate a multi-valued map $\Gamma_c$ with each experimental context $c$. A context $c$ is defined as a particular experimental arrangement of measurement devices, preparation devices etc. Thus, an experiment involving several measurement contexts actually corresponds to several {\em sub-experiments} each associated with a different context. If we combine probabilities conditioned on different contexts it is quite possible to violate the CHSH inequalities. Within any one context $c$, only a subset of possible measurements can actually be performed, and we cannot associate definite relative frequencies with outcomes of measurements that cannot be made within $c$. The multi-valued map $\Gamma_c$ only allows us to associate upper and lower subjective probabilities with UPVs. In this sense counterfactual outcomes are related to measurement contexts through the multi-valued maps.

In the EPRB setup the different contexts are $a,~a',~b,~b',~ab,~a'b,~ab',~a'b'$ corresponding to different possible settings of the SG devices. The contexts $a, a'$ make no reference to the the right particle's SG device, and likewise $b, b'$ do not refer to the left particle's SG device. The remaining contexts (namely $ab,~a'b,~ab'$ and $a'b'$) specify an arrangement of both devices simultaneously. 

Let us consider first a complete table of known values $\Omega \equiv \Omega_{\rm KPV}$ for the outcomes of measurements of the random variables $A, A', B, B' :\Omega \to \{\pm\}$. The table can be divided into rows $\Omega_\nu,~\nu=1,...,N$ and columns $\Omega_c,~c=a,a',b,b'$. The intersection $\Omega_{\nu c}:=\Omega_\nu \cap \Omega_c$ is simply the singleton set $\{\omega_{\nu c}\}$ consisting of the $\nu c$'th table entry. Since every entry has value $+$ or $-$ we can partition $\Omega$ in terms of disjoint subsets as $\Omega=\Omega^+ \cup \Omega^-$. We can now define useful intersections such as $\Omega_{\nu c}^+ := \Omega_{\nu c}\cap \Omega^+$, which is empty if $\omega_{\mu c} = -$ and is equal to $\Omega_{\nu c}$ otherwise.

Considering only a single particle, we can define the following relative frequencies
\begin{align}\label{prob1}
&P(C=\pm) = P(\Omega_c^\pm) = {1\over N}\sum_{\nu =1}^N |\Omega_{\nu c}^+| = {N_c^\pm\over N}
\end{align}
where $C=A,A',B,B'$, $c=a,a',b,b'$. Here $|S|$ is used to denote the cardinality of the set $S$, and $N^+_c = |\Omega_c^+|$ and $N^-_c= |\Omega_c^-|$ denote the number of entries in the $c$'th column with value $+$ and value $-$ respectively; $N_c^+ +N_c^- = N$. Considering both particles together we can define the joint coincidence frequencies
\begin{align}\label{prob2}
P(\Omega^j_c,\Omega^k_{c'}) := {1\over N}\sum_{\nu =1}^N |\Omega_{\nu c}^j| |\Omega_{\nu c'}^k| \equiv {N_{cc'}^{jk} \over N}
\end{align}
where $j=\pm$ and $k=\pm$. The above probabilities can be used to define individual averages and joint correlations as
\begin{align}
&\langle C\rangle = {1\over N}\sum_{j=\pm} j P(\Omega^j_c),\nonumber \\
&\langle CC' \rangle ={1\over N}\sum_{j,k=\pm} jk P(\Omega^j_c,\Omega^k_{c'}).
\end{align} 
In the case that the table is complete $\Omega = \Omega_{\rm KPV}$ the data necessarily obeys all Bell-type inequalities.

Now consider the case in which the table has missing entries corresponding to UPVs. The total space can be partitioned as before as $\Omega=\Omega^+ \cup \Omega^-$, but now we also have the disjoint partitioning $\Omega = \Omega_{\rm KPV}\cup \Omega_{\rm UPV}$. We can also form the intersections $\Omega_{{\rm KPV},\nu c}^j:=\Omega_{\nu c}^j\cap \Omega_{\rm KPV}$ and $\Omega_{{\rm UPV},\nu c}^j:=\Omega_{\nu c}^j\cap \Omega_{\rm UPV}$. The set $\Omega_{{\rm KPV},\nu c}^j$ is empty if $\omega_{\nu c}$ is unknown and is equal to $\Omega_{\nu c}^j$ otherwise. Likewise the set $\Omega_{{\rm UPV},\nu c}^j$ is empty if $\omega_{\nu c}$ is known and is equal to $\Omega_{\nu c}^j$ otherwise.

With each individual context $c$ we associate the multi-valued map $\Gamma_c : \Omega_c= \bigcup_\nu \Omega_{\nu c} \to \Sigma_{\{\pm\}}$ defined by
\begin{align}
\Gamma_c(\omega_c) := \begin{cases}
 + &\mbox{} \omega \in \Omega_{{\rm KPV},c}^+, \\
 - &\mbox{} \omega \in \Omega_{{\rm KPV},c}^-, \\
 \{\pm\} &\mbox{} \omega \in \Omega_{{\rm UPV},c},
       \end{cases}
\end{align}
and with each joint context $cc'$ we associate the multi-valued map $\Gamma_{cc'} : \bigcup_\nu \Omega_{\nu c} \times \Omega_{\nu c'} \to \Sigma_{\{\pm\}\times \{\pm\}}$ defined by
\begin{align}
&\Gamma_{cc'}(\omega_c,\omega_{c'}) :=\nonumber \\ &~  \begin{cases}
 (j,k) &\mbox{} (\omega_c,\omega_{c'}) \in \bigcup_\nu \Omega_{{\rm KPV},\nu c}^j \times \Omega_{{\rm KPV},\nu c'}^k, \\
 \{(j,+),(j,-)\} &\mbox{}(\omega_c,\omega_{c'}) \in \bigcup_\nu \Omega_{{\rm KPV},\nu c}^j\times \Omega_{{\rm UPV},\nu c'}, \\
 \{(+.j),(-,j)\} &\mbox{} (\omega_c,\omega_{c'}) \in \bigcup_\nu \Omega_{{\rm UPV},\nu c}\times \Omega_{{\rm KPV},\nu c'}^j, \\
 \{\pm\}\times\{\pm\} &\mbox{} (\omega_c,\omega_{c'}) \in \bigcup_\nu \Omega_{{\rm UPV},\nu c}\times \Omega_{{\rm UPV},\nu c'}.
\end{cases}
\end{align}
These maps give rise to multiple upper and lower sets. For example, for $j=\pm$
\begin{align}\label{gam1}
&j_{*\Gamma_c} := \{\omega_c : \emptyset \neq \Gamma_c(\omega_c)\subset \{j\} \}= \Omega_{{\rm KPV},c}^j \nonumber \\ &j^{*\Gamma_c} := \{\omega_c : \Gamma_c(\omega_c)\cap\{j\} \neq \emptyset \}= \Omega_{{\rm KPV},c}^j\cup\Omega_{{\rm UPV},c}
\end{align}
and for $j,k=\pm$
\begin{align}\label{gam2}
&(j,k)_{*\Gamma_{cc'}}=\bigcup_\nu \Omega_{{\rm KPV},\nu c}^j \times \Omega_{{\rm KPV},\nu c'}^k\nonumber \\
&(j,k)^{*\Gamma_{cc'}}= \nonumber \\ &\qquad \bigcup_\nu (\Omega_{{\rm KPV},\nu c}^j\cup\Omega_{{\rm UPV},\nu c}) \times (\Omega_{{\rm KPV},\nu c'}^k\cup\Omega_{{\rm UPV},\nu c'}).
\end{align}
We can now define upper and lower probabilities using (\ref{prob1}) and (\ref{prob2}). Since each multi-valued map is associated with a specific context, so too are the upper and lower probabilities. For example, using $\Gamma_c$ we have
\begin{align}
&P_{*c}(j) :=P(j_{*\Gamma_c})= {|\Omega_{{\rm KPV},c}^j|\over N},\nonumber \\ 
&P^{*c}(j) :=P(j^{*\Gamma_c})= {|\Omega_{{\rm KPV},c}^j|+|\Omega_{{\rm UPV},c}|\over N}.
\end{align}
The difference $P^{*c}(\Omega_c^j)-P_{*c}(\Omega_c^j) = P(\Omega_{{\rm UPV},c})$ represents Dempster's ``don't know" probability associated with $\Omega_{{\rm UPV},c}$. That this quantity is nonzero reflects the fact that we cannot reveal any information about the value of the random variable $C$, within the context $c'$. More colloquially, we ``don't know" what the values of $C$ are, if what we are measuring is $C'$.

In the case of measurements on both particles we have within the context $cc'$
\begin{align}
&P_{*cc'}(j,k) :=P[(j,k)_{*\Gamma_c}]= {1\over N}\sum_{\nu =1}^N |\Omega_{{\rm KPV},\nu c}^j| |\Omega_{{\rm KPV},\nu c'}^k|,\nonumber \\
&P^{*cc'}(j,k) :=P[(j,k)^{*\Gamma_c}]\nonumber \\ & ~= {1\over N}\sum_{\nu =1}^N |\Omega_{{\rm KPV},\nu c}^j\cup\Omega_{{\rm UPV},\nu c}||\Omega_{{\rm KPV},\nu c'}^k\cup\Omega_{{\rm UPV},\nu c'}|.
\end{align}
In this case the ``don't know" difference is associated with the cases in which we do not know the value of $C$ for the left-particle {\em or} we do not know the value of $C'$ for the right-particle. If we restrict our attention to a single one-particle context $c$, then we cannot meaningfully associate frequencies with the values of $C'\neq C$. Similarly in the two-particle case restricted to the context $cc'$ we cannot meaningfully associate frequencies with a pair of observables $(D,D')$ for which $D \neq C$ or $D' \neq C'$. We can only meaningfully give subjective upper and lower probability intervals in these cases. Combinations of these subjective probabilities are quite capable of violating Bell-type inequalities, though the sense in which such {\em combinations} are really meaningful raises delicate questions.

There is however, a way to violate the CHSH inequalities with meaningful contextual averages, which are naturally obtained from the Dempster-Shafer multi-valued maps. {\em Each context $c$ defines an observable $C$ whose domain is the set of points for which $\Gamma_c$ is single-valued}. The domain of such an observable is called a {\em domain of certainty} --- a notion that plays a central role in the Dempster-Shafer treatment we employ. Within phenomenological treatments to EPRB-type experiments observables with disjoint domains of certainty could be termed {\em incompatible}. For example, considering only one particle we define $C:\Omega_{{\rm KPV},c} \to \{\pm\}$ by $C:= \Gamma_c|_{\rm KPV}$. In the two-particle case we can also define the product variable $CC':  \bigcup_\nu  \Omega_{{\rm KPV},\nu c}\times \Omega_{{\rm KPV},\nu c'}\to \{{\pm}\}$ by $CC' :=\Gamma_{cc'}^1\Gamma_{cc'}^2|_{\rm KPV}$, which denotes the product of the cartesian components of $\Gamma_{cc'}$ restricted to the subset $\bigcup_\nu \Omega_{{\rm KPV},\nu c}\times \Omega_{{\rm KPV},\nu c'}$. Any two distinct random variables so defined are incompatible. With respect to these variables averages are only defined over domains of certainty;
\begin{align}
&\langle C \rangle_c = \sum_{j=\pm} \sum_{\nu=1}^N {j|\Omega_{{\rm KPV},\nu c}^j|\over |\Omega_{{\rm KPV},c}|},\nonumber \\
&\langle CC' \rangle_{cc'} =\sum_{j,k=\pm} \sum_{\nu=1}^N {jk|\Omega_{{\rm KPV},\nu c}^j||\Omega_{{\rm KPV},\nu c'}^k|\over |\bigcup_\mu \Omega_{{\rm KPV},\mu c}\times \Omega_{{\rm KPV},\mu c'}|}.
\end{align}
Substituting these expressions into (\ref{chsh}) it is clear that (\ref{chsh}) can be violated. In fact, the upper bound on $|f|$ becomes $4$ rather than $2$.

Along with the above averages we can define context restricted probabilities
\begin{align}
&P_c(C=j) := \sum_{\nu=1}^N {|\Omega_{{\rm KPV},\nu c}^j|\over |\Omega_{{\rm KPV},c}|},\nonumber \\
&P_{cc'}(C=j,C'=k) := \sum_{\nu=1}^N {|\Omega_{{\rm KPV},\nu c}^j||\Omega_{{\rm KPV},\nu c'}^k|\over |\bigcup_\mu \Omega_{{\rm KPV},\mu c}\times \Omega_{{\rm KPV},\mu c'}|}.
\end{align}
The first of these represents the frequency with which the outcome $C=j$ is obtained given that the experiment is actually set up to measure $C$, i.e., given that the context is $c$. The second represents the probability that $(j,k)$ is obtained given that both $C$ and $C'$ are actually simultaneously measured, i.e., given that the context is $cc'$. The relevance of this type of conditional probability in relation to EPRB-type experiments was first pointed out by A. Fine \cite{fine_local_1982}. In the treatment above context conditioned probabilities and averages generally violate all Bell-type inequalities, but nothing about this more general {\em probabilistic} treatment precludes the {\em physical} assumptions of classical determinism and locality. 

\section{Including hidden-variables}\label{hv}

Most treatments of EPRB-type experiments including Bell's original treatment, start with the assumption that for each particle pair we can use hidden-variables to give a complete, classically deterministic specification of the real experimental state. These hidden-variables are assumed to belong to a single Kolmogorov probability space. Averages are defined over this one space, and the CHSH inequalities (\ref{chsh}) necessarily hold. In this section we will relax the latter assumption and define contextual random variables whose domains are the domains of certainty of specific Dempster-Shafer multi-valued maps.

\subsection{Rigorous hidden-variable treatment}\label{conven}

First, for comparative purposes, we provide a rigorous formulation of the CHSH inequalities. We consider the standard EPRB setup in which the spins of two particles produced by a common source are measured. We formulate the present treatment within a single Kolmogorov space $(\Omega,\Sigma_\Omega,\mu)$. We define the variables $A,A':\Omega\to \{\pm\}$ representing spin observables in the directions $a,a'$ for the left particle, and similarly we define the spin observables $B,B':\Omega\to \{\pm\}$ for the right particle.
The directions of these spin observables coincide with the contexts $a,a',b,b'$ referring to the SG device alignments. We define the sets $\Omega_c:=C^{-1}(\{\pm\})$ and $+_c:=C^{-1}(+)$ where $C=A,A',B,B'$ and $c=a,a',b,b'$. We also define the real random vector $X:\Omega \to \{\pm\}^4$ by $X(\omega):=(A(\omega),A'(\omega),B(\omega),B'(\omega))$. Finally we assume that $\omega\in \Omega$ can be viewed as giving a complete description of the underlying reality within the experiment, i.e., $\omega$ represents a complete ontic state of the total physical system. With everything defined as such we can now give the two-particle probabilities relevant to Bell-type inequalities as
\begin{align}\label{jps}
\mu(C=j,C'=k) \equiv \int_{j_c\cap k_{c'}} d\mu(\omega).
\end{align}
More generally we have
\begin{align}
\mu\left(X=(j,k,l,m)\right) \equiv \int_{j_a\cap k_{a'}\cap l_b\cap m_{b'}} d\mu(\omega).
\end{align}
There are $2^4=16$ permutations of the outcomes $j,k,l,m=\pm$ appearing in the above expression, giving the same number of probabilities $\mu\left(X=(j,k,l,m)\right)$. These probabilities act as a basis in the sense that they can be used to express any other (absolute) probability. Examples of single-particle and two-particle probabilities are given by
\begin{align}
&\mu(A=j) = \sum_{k,l,m} \mu\left(X=(j,k,l,m)\right), \nonumber \\ &\mu(A=j,B=l) = \sum_{k,m} \mu\left(X=(j,k,l,m)\right).
\end{align}
Single-particle and two-particle averages are given by
\begin{align}\label{hvavs}
&\langle C \rangle := \int_\Omega d\mu(\omega) \, C(\omega), \nonumber\\  &\langle CC' \rangle := \int_\Omega d\mu(\omega) \, C(\omega)C'(\omega).
\end{align}
Despite the fact that for a single particle we cannot simultaneously attribute known possessed values to both of the observables $A$ and $A'$, equation (\ref{hvavs}) does not distinguish between the averages like $\langle AA'\rangle$, and meaningful two-particle correlations such as $\langle AB\rangle$. Similarly the formalism does not itself distinguish between the probabilities $\mu\left(X=(j,k,l,m)\right)$ in which outcomes are simultaneously associated with all observables, and probabilities like $\mu(A=j,B=l)$, which only associates outcomes with observables that can be simultaneously measured. In other words, because it has been built using a single Kolmogorov space the above hidden-variable approach treats counterfactual outcomes in the same way as the conventional phenomenological approaches discussed in section \ref{phen}. In short, the above approach assumes that counterfactual outcomes can be treated in the same way as factual outcomes. It is convenient to refer to this assumption as simply {\em the counterfactual assumption}.

The quantities above can be written in terms of concrete probability densities by defining a Borel measurable chart $\Lambda : \Omega \to {\mathbb R}^n$ consisting of hidden-variables that reveal the underlying physical states $\omega\in \Omega$. If we denote by ${\mathcal B}({\mathbb R}^n)$ the Borel subsets of ${\mathbb R}^n$, the triple $({\mathbb R}^n,{\mathcal B}({\mathbb R}^n),P_\Lambda :=\mu \circ \Lambda^{-1})$ defines a concrete Kolmogorov space. One can then define the observables ${\tilde C} := C\circ\Lambda^{-1}$ over this space, and one can define the joint distribution function $F_\Lambda : {\mathbb R}^n \to [0,1]$ by $F_\Lambda(\lambda_1,...,\lambda_n) :=P_\Lambda(\Lambda_1 \leq \lambda_1,...,\Lambda_n\leq \lambda_n)$. The {\em probability density} associated with $P_\Lambda$ is defined by $\rho_\Lambda := \partial^n F_\Lambda/ \partial \lambda_1...\partial \lambda_n$. The number $\rho_\Lambda(\lambda)$ characterises the observer's epistemic, but purely ``Kolmogorovian" uncertainty, as to the underlying physical state $\omega,~\Lambda(\omega)=\lambda$. This interpretation is inherited from the interpretation given to the Kolmogorov probability measure $\mu$. The assumption of classical determinism means the observable ${\tilde C}$ is assumed to {\em possess} the definite value ${\tilde C}(\lambda)$ in the state $\lambda$. Note that the formalism distinguishes between two distinct types of state---an ontic state $\omega$ (or equivalently $\lambda=\Lambda(\omega)$), and an epistemic state $\rho_\Lambda$. A simple example of the above formalism at work is given later on in section \ref{rocon}. For now we note that equations (\ref{jps}) and (\ref{hvavs}) can be written
\begin{align}\label{jps2}
P_\Lambda({\tilde C}=j,{\tilde C}'=k) = \int_{{\tilde j}_c\cap {\tilde k}_{c'}} dP_\Lambda(\lambda) = \int_{{\tilde j}_c\cap {\tilde k}_{c'}} d^n\lambda\, \rho_\Lambda(\lambda)
\end{align}
and
\begin{align}\label{hvavs2}
&\langle {\tilde C} \rangle = \int_{{\mathbb R}^n} d^n\lambda \, \rho_\Lambda(\lambda){\tilde C}(\lambda) \nonumber \\
&\langle {\tilde C}{\tilde C}' \rangle = \int_{{\mathbb R}^n} d^n\lambda \, \rho_\Lambda(\lambda){\tilde C}(\lambda){\tilde C}'(\lambda)
\end{align}
respectively. Within the framework discussed above the CHSH inequalities (\ref{chsh}) necessarily hold.

\subsection{A contextual Dempster-Shafer approach}\label{ds}

We now offer an alternative approach to that above for the modeling of EPRB-type experiments. As in section \ref{phen} we achieve this using multi-valued maps. The resulting framework shares many features in common with Khrennikov's contextual framework \cite{khrennikov_contextual_2009}. Our starting point is the idea that even if we only considered measuring the spin of a single particle in the different directions $c=a,a'$, a single Kolmogorov probability space $(\Omega,\Sigma_\Omega,\mu)$ would not offer an adequate description of the experiment being envisioned. Using only a single space makes it impossible to account for the fact that different spin directions are not simultaneously measured. The latter is an operational fact, which must be properly encoded within any theoretical treatment, regardless of whether or not we make particular physical assumptions like determinism or locality. Thus, the theory based on a single Kolmogorov space must be appended in order to properly account for the incompatibility between different experimental contexts. In short, we avoid making the counterfactual assumption.

A single-particle context is assumed to be determined by the alignment of the SG device. As both Bohr and later Bell repeatedly emphasized should be the case, the model of an experiment should describe the entire experiment, rather than just the systems being measured \cite{bell_speakable_2004}. With this in mind we define $\Omega = \bigcup_c \Omega_c$ where $c=a,a',b,b'$ are distinct SG device contexts. Each $\Omega_c$ can be decomposed as $\Omega_c =+_c\cup-_c$ with the sets $\pm_c$ corresponding to the possible spin outcomes along the direction $c$.

Just like in section \ref{eprb1} we associate with each $c$ a multi-valued map $\Gamma_c:\Omega \to \Sigma_{\{\pm\}}$ defined by
\begin{align}
\Gamma_c(\omega):=\begin{cases}
 + &\mbox{} \omega \in +_c, \\
 - &\mbox{} \omega \in -_c, \\
 \{\pm\} &\mbox{} \omega \not\in \Omega_c.
       \end{cases}
\end{align}
The uncertainty in the event space within the context $c$, is represented by the set ${\overline \Omega}_c$ which denotes the complement of $\Omega_c$. We can also define the joint-context multi-valued maps $\Gamma_{cc'}:\Omega \to \Sigma_{\{\pm\}}$ by
\begin{align}
\Gamma_{cc'}(\omega):=\begin{cases}
jk &\mbox{} \omega \in j_c\cap k_{c'}, \\
\{\pm\} &\mbox{} \omega \not\in {\Omega}_c\cap\Omega_{c'}. \\
\end{cases}
\end{align}
As before we define the random variables $\Gamma_c|_{\Omega_c}=:C:\Omega_c\to \{\pm\}$, whose domains are the domains of certainty of the $\Gamma_c$. We also define the product variables $CC':=\Gamma_{cc'}|_{\Omega_c\cap\Omega_{c'}}$. 
With respect to these observables single-particle and two-particle probabilities are defined by
\begin{align}\label{p}
&\mu(j_c|\Omega_c) := {1\over \mu(\Omega_c)}\int_{j_c} d\mu(\omega), \nonumber \\ 
&\mu(j_c\cap k_{c'}|\Omega_c\cap\Omega_{c'}) :=  {1\over \mu(\Omega_c\cap\Omega_{c'})} \int_{j_c\cap k_{c'}} d\mu(\omega).
\end{align}
The expressions in (\ref{con1}) are normalised over the domains of certainty defined by $\Gamma_c$ and $\Gamma_{cc'}$ respectively, and can be interpreted as conditional probabilities. In a similar fashion single-particle averages and two particle correlations are defined as
\begin{align}\label{con1}
&\langle C \rangle_c := {1\over \mu(\Omega_c)} \int_{\Omega_c} d\mu(\omega) \,C(\omega), \nonumber \\ 
&\langle CC' \rangle_{cc'} :=  {1\over \mu(\Omega_c\cap\Omega_{c'})} \int_{\Omega_c\cap\Omega_{c'}} d\mu(\omega) \,C(\omega)C'(\omega)
\end{align}
where again each expression is normalised over a context specific domain of certainty, and so represents a conditional expectation. If we substitute correlations of the form given in (\ref{con1}) into (\ref{chsh}) we obtain as in section \ref{eprb1} an upper bound on $|f|$ of $4$ rather than $2$. Thus, conditional probabilities and correlations do not satisfy Bell-type inequalities. Of course, the interpretation of quantum probabilities pertaining to EPRB experiments as classical conditional probabilities is well-known \cite{redei_john_2001,khrennikov_classical_2014}.

As in section \ref{conven} we can map $\Omega$ into ${\mathbb R}^n$. However, since in general we expect the epistemic states of an observer to depend on the context, we now associate with each context $c$ a ``local" (as opposed to global) chart $\Lambda_c :\Omega_c\to {\mathbb R}^n$ defined over the domain of certainty $\Omega_c$. With each joint context $cc'$ we associate another chart $\Lambda_{cc'}:\Omega_c\cap\Omega_{c'}\to {\tilde \Omega}_c\cap {\tilde \Omega}_{c'}$ defined over the domain of certainty $\Omega_c \cap \Omega_{c'}$. The local charts $\Lambda_c$ give rise to measures $P_c$, epistemic states $\rho_c$, and observables ${\tilde C}$ defined over ${\tilde \Omega}_c:=\Lambda_c(\Omega_c) \in {\mathcal B}({\mathbb R}^n)$. Analogous measures, states and observables are associated with the joint contexts $cc'$. The set ${\tilde \Omega}_c$ is the ``concrete" domain of certainty in ${\mathcal B}({\mathbb R}^n)$, that corresponds to the ``abstract" domain of certainty $\Omega_c \in \Sigma_\Omega$. The probabilities in (\ref{p}) can now be written
\begin{align}\label{p2}
&p_c(C=j):=P_c({\tilde j}_c|{\tilde \Omega}_c) := {1\over P_c({\tilde \Omega}_c)}\int_{{\tilde j}_c} d^n\lambda\,\rho_c(\lambda), \nonumber \\ 
&p_{cc'}(C=j,C'=k):=P_{cc'}({\tilde j}_c\cap {\tilde k}_{c'}|{\tilde \Omega}_c\cap{\tilde \Omega}_{c'}) \nonumber \\ &\qquad \qquad ~~ \, :=  {1\over P_{cc'}({\tilde \Omega}_c\cap{\tilde \Omega}_{c'})} \int_{{\tilde j}_c\cap {\tilde k}_{c'}} d^n\lambda\,\rho_{cc'}(\lambda).
\end{align}
The notation $P(\cdot | {\tilde \Omega}_c)$ does not necessarily denote a conventional conditional probability within ${\mathbb R}^n$, rather it indicates that all probabilities pertaining to context $c$ are normalised over ${\tilde \Omega}_c$. In general one need not require that $P_c({\tilde \Omega}_c)\leq1$. The normalisation factors in (\ref{p2}) can be absorbed into the definition of the densities to yield absolute probabilities over the domains of certainty. More precisely, defining $P'_c:=P_c/P_c({\tilde \Omega}_c)$ and $P'_{cc'}:=P_{cc'}/P_{cc'}({\tilde \Omega}_c\cap{\tilde \Omega}_{c'})$, we have
\begin{align}\label{p2b}
&p_c(C=j)=P'_c({\tilde j}_c)=\int_{{\tilde j}_c} d^n\lambda\,\rho'_c(\lambda),\nonumber \\
&p_{cc'}(C=j,C'=k) =P'_{cc'}({\tilde j}_c\cap {\tilde k}_{c'})= \int_{{\tilde j}_c\cap {\tilde k}_{c'}} d^n\lambda\,\rho'_{cc'}(\lambda).
\end{align}
These normalised measures allow one to define conditional probabilities within the domains of certainty in the usual way. The averages in (\ref{con1}) can be written
\begin{align}\label{con2}
&\langle C\rangle_c = {1\over P_c({\tilde \Omega}_c)}\int_{{\tilde \Omega}_c} d^n\lambda \, \rho_c(\lambda)\,{\tilde C}(\lambda),\nonumber \\
&\langle CC'\rangle_{cc'} = {1\over P_c({\tilde \Omega}_c\cap{\tilde \Omega}_{c'})}\int_{{\tilde \Omega}_c \cap {\tilde \Omega}_c'} d^n\lambda\, \rho_c(\lambda)\,{\tilde C}(\lambda){\tilde C}'(\lambda).
\end{align}
Whenever the joint correlations in (\ref{con2}) can be expressed in the form given in (\ref{hvavs2}), the CHSH inequalities in (\ref{chsh}) hold. This requires the existence of a common density $\rho(\lambda)$ corresponding to a global chart $\Lambda:\Omega\to {\mathbb R}^n$, such that every probability is normalised over the total space ${\tilde \Omega}$. Such a description would only generally be appropriate if all measurements within the experiment were performed under the same physical conditions, that is, within the same context. Note that when a global chart is used but the probabilities are normalised only over the domains of certainty, they can be interpreted as conditional probabilities as in (\ref{p}).

\subsection{Discussion}

Two different alignments of a SG device produce altogether different inhomogeneous magnetic fields over the spacetime region within which the spin measurement is performed. In general, the physical conditions under which a measurement is performed are at least partly determined by the state of the measuring device, which therefore influences the observed physical events. As Bell himself puts it: ``{\em the results have to be regarded as the joint product of `system' and `apparatus', the complete experimental setup}" \cite{bell_speakable_2004}. In a deterministic theory the set of all possessed values for the observables of a physical system can be divided into KPVs and UPVs. The KPVs are what Bell calls ``results", while obviously the UPVs are not ``results". As such UPVs should {\em not} be regarded as the joint product of system and apparatus. Rather, UPVs depend on the system alone. Moreover, in EPRB-type experiments the UPVs pertain to spin observables that are operationally incompatible with the spin observables that are actually measured to give the KPVs. Crucially therefore, {\em we cannot assume that the set of KPVs is representative of the set of UPVs.}. In other words, if, in determining measurement outcomes, measurement devices are {\em active} rather than {\em passive}, then the counterfactual assumption is not valid whenever operationally incompatible observables are being considered. Thus, we see that Bell's own conception of measurement cannot generally be reconciled with the counterfactual assumption crucial for proving Bell-type inequalities.

In our approach distinct and incompatible states of a measuring device result in different physical contexts $c$. Each context $c$ is associated with a multi-valued map $\Gamma_c$. With respect to a given context $c$, the uncertainty associated with counterfactual events taking place in a different context $c'\neq c$, is non-Kolmogorovian epistemic uncertainty pertaining to the underlying event space. This leads to the conditioning of densities on measurement contexts, and results in probabilities as in (\ref{p}). These probabilities are defined in terms of different measures and are normalised over different domains. As a result they are not constrained by Bell-type inequalities. The contextual framework given here, offers an explanation as to why violations of Bell-type inequalities have actually been observed in the lab \cite{aspect_experimental_1982} without resorting to nonlocality or indeterminism.

\subsubsection{Free will}

In mathematical terms the contextual formulation above uses multiple Kolmogorov probability spaces to model the EPRB experiment. This is deemed appropriate because the experiment actually consists of four incompatible experimental contexts. Thus, rather than using a single measure $P$ corresponding to which there is a single density $\rho$, we use multiple densities $\rho_{c}$, which are labeled by context. The assumption that a single density suffices, so that
\begin{align}\label{freewill}
\rho(\lambda)=\rho_c(\lambda)
\end{align}
is called {\em measurement independence} in \cite{hall_local_2010,barrett_how_2011}, and is discussed in \cite{_exchange_1985}. It is argued that this assumption encodes the ``free will" of the experimenter in choosing measurement settings. Thus, the use of different Kolmogorov probability spaces to model different and incompatible (sub)experiments is somehow seen as violating the assumption that observers have ``free will".

First we note that even when equation (\ref{freewill}) holds, Bell-type inequalities will generally be violated if the probabilities involved are interpreted as conditional probabilities \cite{redei_john_2001,khrennikov_classical_2014,khrennikov_unconditional_2015}. Furthermore, absolute relative frequency probabilities are necessarily absolute Kolmogorovian probabilities and therefore they must satisfy Bell-type inequalities. This means that quantum probabilities violating Bell-type inequalities cannot be interpreted as absolute frequency probabilities \cite{redei_john_2001,khrennikov_unconditional_2015}. A violation of Bell-type inequalities can only be explained through some sort of conditional probability model \cite{redei_john_2001,khrennikov_unconditional_2015}. The challenge one is faced with in constructing a deterministic hidden-variable model, is the reproduction of the quantum predictions using either conventional event-conditioned classical probabilities, or perhaps using context-conditioned classical probabilities \footnote{Context-conditioning is different to event conditioning in that it makes use of multiple probability measures, each associated with a different context. Event-conditioning is the standard conditioning that occurs within a single Kolmogorov space, and is defined using Bayes' rule (see \cite{khrennikov_contextual_2009}).}. Regarding the latter, one can simply interpret ``measurement dependence" as the assumption that incompatible experimental contexts require separate probability spaces. This is justified through the assumption that experimental results are the product of the system and the apparatus. 

Consider, for example, a situation in which a physical system $S$ is prepared at time $-t_0$ and that this results in some ontic (possibly unknown) state $\lambda(-t_0)$. Suppose that at time $t=0$ the system is to be measured for some duration $T$, with a device $M$ whose state $\lambda_c$ is always known and can be controlled by the experimenter. In many cases a reasonable assumption might be that the state of the macroscopic measuring device is stationary $\lambda_c(t)=\lambda_c$.

We can now define a ``free will" condition as the assumption that $\lambda_c$ and $\lambda$ are independent variables for $t<0$. This means that $\lambda_c$ which can be chosen at any time $t<0$ is not constrained by the system $S$. However, during the measurement process $S$ and $M$ will certainly interact, so there is no good reason to assume that $\lambda(t)$ will be independent of $\lambda_c$ for $t\geq 0$. The probability density $\rho(\lambda(t))$ over ontic states taken at the time of the measurement therefore depends on $\lambda_c$ (as well as $\lambda(-t_0)$ and other initial data). We can therefore write $\rho(\lambda(t))=\rho(\lambda(-t_0),\lambda_c,t) \equiv \rho_c(\lambda(-t_0),t)$ for $t\geq 0$. Thus, if the probability to obtain a particular measurement result is calculated from a density over ontic states taken {\em at the time of the measurement}, then the probabilities will be contextual. This does not violate ``free will", but simply indicates that the system and measurement device have interacted, so the result is not ``measurement independent".

The above qualitative analysis implies that the measuring device generally disturbs the system being measured and so it is not passive within the experiment. The assumption of a passive measurement device may often be made when dealing with macroscopic systems in classical physics, but this assumption is not a fundamental postulate of the latter, and it seems ill-justified when considering microscopic systems. Indeed, the assumption that measurement devices are always passive in classical physics introduces a kind of classical measurement problem in which measurement devices are given a privileged role as special systems that do not disturb the systems they interact with. 

\section{Relation to conventional approaches}\label{c2}

In this section we attempt to assess how the rigorous treatment in \ref{conven}, the contextual treatment in \ref{ds}, and conventional hidden-variable treatments found throughout the physics literature, are each related to one another. Throughout the physics literature pertaining to Bell-type inequalities, one frequently encounters heuristic expressions for what is known as {\em Bell's locality assumption} written in the form
\begin{align}\label{heur}
p_\lambda(j,k|a,b) = p^1_\lambda(j|a)p^2_\lambda(k|b).
\end{align}
Here $j$ and $k$ are left and right-particle spin outcomes respectively, $a$ and $b$ are left and right-SG device settings respectively, and as usual $\lambda$ is supposed to offer a complete description of the underlying state of affairs. The labels $1$ and $2$ in (\ref{heur}) refer to the left and right-particles respectively. This particular notation was borrowed from \cite{fry_experimental_2009}. Sometimes one sees $\lambda$ appearing as a conditioning event as in the following expression, borrowed from a recent article \cite{zukowski_quantum_2014}
\begin{align}\label{heur2}
p(j,k|a,b) =  \int d\lambda \, \rho(\lambda)p(j,k|a,b,\lambda).
\end{align}
In our notation (\ref{heur}) would read
\begin{align}\label{ours}
p_{ab}(A=j,B=k|\lambda) = p_a(A=j|\lambda)p_b(B=k|\lambda)
\end{align}
though we have not yet offered an interpretation of the conditioning of our probabilities on an underlying physical state $\lambda$. This will be done in what follows.

\subsection{The role of conditional probabilities}\label{rocon}

The probabilities on the left-hand-side in (\ref{heur2}) are our primary interest as these are the probabilities appearing in Bell-type inequalities. Let us therefore focus on the left-hand-side of (\ref{heur2}). Suppose that we treat the device settings $a$ and $b$ as genuine conditioning events in an event space $\Omega$. This, after all, is what the notations in (\ref{heur}) and (\ref{heur2}) suggest we should do. As soon as we choose to do this it becomes impossible to avoid including a description of the SG devices within the probabilistic treatment of the experiment. Proceeding along this line of inquiry, our aim in this section is to relate the expressions in (\ref{heur}) and (\ref{heur2}) to the expressions derived in \ref{ds}. To this end we interpret the left-hand-side of (\ref{heur2}) as being given by (\ref{p2});
\begin{align}\label{gc}
p_{ab}(A=j,B=k) &=P_{ab}({\tilde j}_a\cap {\tilde k}_b|{\tilde \Omega}_a\cap{\tilde \Omega}_b). 
\end{align}
Now we would like to determine whether or not the right-hand-side of (\ref{heur2}) can be understood as being equal to the right-hand-side of (\ref{gc}). The probabilities in (\ref{heur}) and (\ref{heur2}) are supposed to be conditioned on the settings of the SG devices. This gives the impression that the states of the measuring devices have been properly taken into account. However, the expressions obtained do not represent true conditional probabilities, because they are not normalised as such. Rather they are normalised by $\mu(\Omega)=1$. Perhaps the conditioning is supposed to have been taken care of by the hidden-variables $\lambda$. Let us investigate this possibility.

In the rigorous measure-theoretic formulation of Kolmogorov probability theory (c.f. \cite{gatti_probability_2004,rao_probability_2006,klenke_probability_2013}) a probability density is associated with a real random vector $\Lambda :\Omega\to {\mathbb R}^n$, which defines a measure $P_\Lambda := \mu\circ \Lambda^{-1}:{\mathcal B}({\mathbb R}^n)\to [0,1]$  where ${\mathcal B}({\mathbb R}^n)$ is the set of Borel subsets of ${\mathbb R}^n$. Note that the domain of $P_\Lambda$ implies that $\Lambda^{-1}:{\mathcal B}({\mathbb R}^n)\to \Sigma_\Omega$ denotes the pre-image map, which is well-defined whether or not $\Lambda$ is invertible. In turn one can define an absolutely continuous distribution $F_\Lambda(\lambda) :=P_\Lambda(B_\lambda) \equiv \mu(\Lambda\leq \lambda)$ where $B_\lambda:=(-\infty,\lambda_1]\times...\times(-\infty,\lambda_n]\in{\mathcal B}({\mathbb R}^n)$. The associated density is defined as $\rho_\Lambda:= \partial^n F_\Lambda / \partial\lambda_1...\partial\lambda_n$. Formally $\rho_\Lambda(\lambda)$ can be thought of as the probability that $\Lambda=\lambda$, and could be denoted $\mu(\Lambda=\lambda)$. One can extend the construction to deal with conditional probabilities by defining for $S =\Lambda^{-1}(B) \in \Sigma_\Omega$ the measure $\mu_S :=\mu(\cdot |S) := \mu(\cdot \cap S)/\mu(S)$. With this one can define a measure $P_{\Lambda|B}$ over ${\mathcal B}({\mathbb R}^n)$ as
\begin{align}
P_{\Lambda|B}(B') &:= \mu(\Lambda^{-1}(B')|\Lambda^{-1}(B)) = {1\over \mu(S)}\int_{\Lambda^{-1}(B\cap B')}\hspace*{-0.3cm}d\mu\nonumber \\
&= {1\over P_\Lambda(B)}\int_{B\cap B'} dP_\Lambda.
\end{align}
Then one defines the conditional distribution $F_{\Lambda|B}(\lambda):=P_{\Lambda|B}(B_\lambda)$ and the associated density
\begin{align}\label{rhocon}
\rho_{\Lambda|B}(\lambda):={\partial^n F_{\Lambda|B}(\lambda) \over \partial \lambda_1...\partial\lambda_n} = {\rho_\Lambda(\lambda)\delta_B(\lambda)\over P_\Lambda(B)}={\rho_\Lambda(\lambda)\delta_B \over \int_B d^n\lambda\, \rho_\Lambda(\lambda)} 
\end{align}
where $\delta_B$ is the Dirac measure associated with $B$;
\begin{align}
\delta_B(\lambda)= \begin{cases}
1 &\mbox{} \lambda \in B, \\
0 &\mbox{} \lambda \not\in B. \\
\end{cases}
\end{align}
According to this definition $\rho_{\Lambda|B}(\lambda)$ is normalised to unity indicating that it is a genuine probability density. Equation (\ref{rhocon}) yields for a subset $S'=\Lambda^{-1}(B')\in \Sigma_\Omega$ with $B'\in{\mathcal B}({\mathbb R})$, the following expression for a conditional probability written in terms of the corresponding density;
\begin{align}
\mu(S'|S)&=P_\Lambda(B'|B)={P_\Lambda(B'\cap B) \over P_\Lambda(B)}=\int_{B'}d^n\lambda\, \rho_{\Lambda|B}(\lambda) \nonumber \\ &= {\int_{B'\cap B}d^n\lambda\, \rho_\Lambda(\lambda) \over \int_B d^n\lambda\, \rho_\Lambda(\lambda)}.
\end{align}
Finally, we can define the conditional probability $P_\Lambda(B|\lambda)$ by
\begin{align}\label{pcon}
P_\Lambda(B|\lambda)\rho_\Lambda(\lambda) = P_\Lambda(B)\rho_{\Lambda|B}(\lambda)
\end{align}
whenever $\rho_\Lambda(\lambda) \neq 0$ almost everywhere (except on a set of measure zero). Comparing this expression with (\ref{rhocon}) yields
\begin{align}\label{pcondelta}
P_\Lambda(B|\lambda) = \delta_B(\lambda).
\end{align}
This result is precisely what we should expect to find, because we are assuming that $\lambda$ represents the precise state of affairs within the underlying physical reality. If we are given $\lambda$ we can be certain which events will and will not occur. The quantity $P_\Lambda(B|\lambda)$ represents the probability of event $B$ {\em given} $\lambda$. Thus, $P_\Lambda(B|\lambda)$ must be either $0$ or $1$. More precisely, the event $B$ is certain to occur if $\lambda\in B$ and certain not to occur otherwise, hence $P_\Lambda(B|\lambda) = \delta_B(\lambda)$. Another way to view $P_\Lambda(B|\lambda)$ is as the probability of $B$ given that the epistemic state $\rho_\Lambda$ is the delta function $\rho_\Lambda(\lambda')=\delta(\lambda-\lambda')$, which corresponds to the situation in which we have complete knowledge of the underlying reality. Borrowing terminology from quantum theory, delta-function epistemic states $\rho_\Lambda(\lambda')=\delta(\lambda-\lambda')$ could be termed {\em pure states}, with all other epistemic states termed {\em mixed states}. These pure states are clearly in one-to-one correspondence with the ontic states $\lambda$. Now, from either (\ref{pcondelta}), or directly from (\ref{pcon}), it follows that
\begin{align}\label{pbf}
P_\Lambda(B) = \int d^n\lambda\, \rho_\Lambda(\lambda)P_\Lambda(B|\lambda) =  \int_B d^n\lambda\, \rho_\Lambda(\lambda)
\end{align}
of which the first equality appears to be quite close to what we see in (\ref{heur2}). Before we use the above formalism in analysing the conventional hidden-variable approaches, it may be instructive to see it in action using a simple example.

Consider a point particle moving in one-dimensional Euclidean space $E^1$. With initial conditions the second order dynamics resulting from Newton's laws defines a well-posed Cauchy problem. A position value and velocity (momentum) value suffice to give a complete physical description of the particle. Thus, the state (event) space is $\Omega =T^*E^1$, which denotes the cotangent bundle of $E^1$. We can turn the manifold $T^*E^1$ into a Kolmogorov space by equipping it with a Kolmogorov measure $\mu:\Sigma_{T^*E^^1}\to[0,1]$ where $\Sigma_{T^*E^1}$ is a suitable $\sigma$-algebra. Since $T^*E^1$ is a flat manifold it admits a global coordinate chart $\Lambda: T^*E^1 \to {\mathbb R}^2$, which is associated with some family of observers ${\mathcal O}$. A particle state is a phase point $\omega\in T^*E^1$ with coordinate representation $\Lambda(\omega) =(x,p)\in {\mathbb R}^2$ relative to ${\mathcal O}$. A particle observable is a suitably well-behaved (e.g. smooth, square-integrable) function $f:T^*E^1\to {\mathbb R}$, which admits the coordinate representation ${\tilde f}:=f\circ \Lambda^{-1}:{\mathbb R}^2\to {\mathbb R}$. Finally $P_\Lambda:=\mu\circ\Lambda^{-1}$ defines a normalised Kolmogorov measure on ${\mathbb R}^2$ and an epistemic state associated with the observers ${\mathcal O}$ is an integrable density $\rho_\Lambda$ such that
\begin{align}
P_\Lambda({\tilde U}) =\int_{\tilde U}dx \, dp \,\rho_\Lambda(x,p)
\end{align}
where ${\tilde U}:=\Lambda(U)$ and $U\in \Sigma_{T^*E^1}$. Different choices of $\rho_\Lambda$ correspond to different choices of $P_\Lambda$, which correspond to different choices of $\mu$. This means that different choices of $\rho_\Lambda$ encode different epistemic states of Kolmogorovian uncertainty regarding the ontic state of the particle $\omega\in T^*E^1$. The average value of observable $f$ measured by observers ${\mathcal O}$ with epistemic state $\rho_\Lambda$ is
\begin{align}
\langle f \rangle := \int_{{\mathbb R}^2} dx \, dp \, \rho_\Lambda(x,p){\tilde f}(x,p).
\end{align}
The probability that relative to ${\mathcal O}$ the particle's state belongs in ${\tilde U}:=\Lambda(U)$ is
\begin{align}\label{pu}
\mu(U)=P_\Lambda({\tilde U}) &=\int_{\tilde U} dx\, dp\, \rho_\Lambda(x,p) \nonumber \\ &=\int_{{\mathbb R}^2} dx\, dp\, \delta_{\tilde U}(x,p) \rho_\Lambda(x,p).
\end{align}
If $\rho_\Lambda(x,p)=\delta(x-x_0)\delta(p-p_0)$ the observers know with certainty that the state of the particle has coordinate representation $(x_0,p_0)$ relative to them. They therefore know with certainty whether or not the state belongs in $U$, i.e., whether or not $(x_0,p_0)$ belongs in ${\tilde U}$. For this particular epistemic state the probability in (\ref{pu}) is equal to one if $(x_0,p_0)\in {\tilde U}$ and equal to zero otherwise, i.e.,
\begin{align}\label{pu2}
P_\Lambda({\tilde U})&= \int_{{\mathbb R}^2} dx\, dp\, \delta_{\tilde U}(x,p) \delta(x-x_0)\delta(p-p_0) \nonumber \\&= \delta_{\tilde U}(x_0,p_0) \equiv P_\Lambda({\tilde U}|x_0,p_0).
\end{align}
For general $\rho_\Lambda$, (\ref{pu}) can be written
\begin{align}
P_\Lambda({\tilde U})&= \int_{{\mathbb R}^2} dx\, dp\, \rho_\Lambda(x,p)\delta_{\tilde U}(x,p) \nonumber \\&= \int_{{\mathbb R}^2}  dx\, dp\,\rho_\Lambda(x,p)P({\tilde U}|x,p)
\end{align}
which is just a specific example of expression (\ref{pbf}).

\subsection{Comparison of conventional and contextual approaches}

Having determined with the formalism above precisely how to interpret probabilities that are conditioned on the ontic state $\lambda$ and having seen this formalism work in a simple setting, we now have all the ingredients we need in order to compare conventional approaches with our contextual Dempster-Shafer approach. First we use (\ref{pcondelta}) to interpret the probabilities in (\ref{ours}) as
\begin{align}\label{w}
p_{ab}(A=j,B=k|\lambda)= \delta_{{\tilde j}_a\cap{\tilde k}_b}(\lambda),
\end{align}
which using (\ref{p2b}) and (\ref{pbf}) yields
\begin{align}\label{o}
p_{ab}(A=j,B=k) = \int d^n\lambda \, \rho'_{ab}(\lambda)p_{ab}(A=j,B=k|\lambda).
\end{align}
Equation (\ref{o}) is similar to (\ref{heur2}), but of course the probabilities in (\ref{o}) are defined differently to those in (\ref{heur2}). Unlike those in (\ref{heur2}), the probabilities in (\ref{o}) do not obey Bell-type inequalities. The main difference between (\ref{o}) and (\ref{heur2}) is that in the latter the labels $a,~b$ pertaining to the measurement devices appear to be merely notational tools, while in the former they take an active role in defining the probability measures $P_{ab}$ (and hence $\rho_{ab}$), and in properly normalising the resulting probabilities.

The question arises as to the sense in which our approach can be considered local. Although Bell considered his locality assumption (\ref{heur}) crucial for the proof of his theorem, it was noted some time ago by Accardi, that the crucial assumption underlying Bell-type inequalities is the {\em counterfactual assumption} not Bell's {\em locality assumption} \cite{accardi_locality_2000,accardi_epr_2008}. In fact Accardi has shown that Bell's inequalities hold for spin random-variables defined over a single Kolmogorov space, even if one assumes the negation of his locality assumption \cite{accardi_locality_2000}.

The idea underlying our contextual framework is that observed physical events in an experiment are dependent on both the measured systems and the measurement devices. While this invalidates the counterfactual assumption, Bell's locality assumption (\ref{ours}) remains perfectly intact, and is satisfied as an immediate consequence of (\ref{w}) and the identity $\delta_{A\cap B}\equiv\delta_A\delta_B$. Moreover, given that $\lambda$ is supposed represent the complete ontic state of the system, it is difficult to see how to interpret probabilities conditioned on $\lambda$ in any other way besides as Dirac measures of the form $\delta_A(\lambda)$. Indeed the formalism of section \ref{c2}, which leads to (\ref{pcondelta}), is general enough to include any ontic state of the type encountered in classical physics. But if this is indeed our only option, then $\lambda$-conditioned probabilities will always trivially satisfy Bell's locality assumption.


\section{Example}\label{eg}

In this section we examine a model of EPRB-type experiments that fits within our contextual approach. The model is quite similar to a model originally proposed by Barut and Meystre \cite{barut_classical_1984}, and is designed to reproduce the quantum predictions.

\subsection{The model}

To begin with we consider one particle and two orientations $c=a,a'$ of an SG device. For simplicity we restrict the model to two spatial dimensions (the $xz$-plane). The spin observable in the $c$-direction $C=:\Omega_c \to \{\pm\}$ is defined by
\begin{align}\label{C}
C(\omega_c) = \begin{cases}
+ & \mbox{} \omega_c \in +_c \\
-  & \mbox{} \omega_c \in -_c.
\end{cases}
\end{align}
We consider a classical particle with angular-momentum spin unit-vector ${\bf S}$, which can be fully specified by an angle $\lambda$ relative to the $z$-axis. We fix the unit vector ${\bf c}={\bf a},{\bf a}'$ giving the alignment of the SG device, which can be fully specified by an angle $\lambda_{\bf c}$. We define the chart $\Lambda_c:\Omega_c \to {\mathbb R}$ by
\begin{align}\label{LC}
\Lambda_c(\omega_c) = \begin{cases}
\lambda -\lambda_{\bf c} & \mbox{} \omega_c \in +_c \\
\lambda - \lambda_{-{\bf c}} & \mbox{} \omega_c \in -_c.
\end{cases}
\end{align}
We now choose the sets $\pm_c$ such that together (\ref{C}) and (\ref{LC}) yield the representation ${\tilde C}:{\tilde \Omega}_c \to\{\pm\}$ defined by
\begin{align}
{\tilde C}(\lambda)=\begin{cases}
+ & \mbox{} \lambda -\lambda_{\bf c} \in \left[{-\pi/2}, {\pi/2}\right) \\
- & \mbox{} \lambda -\lambda_{-\bf c} \in \left[{-\pi/2}, {\pi/2}\right).
\end{cases}
\end{align}
Since $\lambda_{-{\bf c}} \equiv \lambda_{\bf c} +\pi$, the sets ${\tilde \pm}_c$ are given by
\begin{align}\label{til}
{\tilde +}_c = \left[\lambda_{\bf c}-{\pi\over 2},\lambda_{\bf c}+{\pi\over 2}\right),\qquad
{\tilde -}_c = \left[\lambda_{\bf c}+{\pi\over 2},\lambda_{\bf c}+{3\pi\over 2}\right).
\end{align}
which respectively represent the north and south hemispheres of the sphere for which ${\bf c}$ specifies the north pole. The set ${\tilde \Omega}_c = {\tilde +}_c\cup{\tilde -}_c$ evidently represents to the whole sphere. 

The density associated with the measure $P_c$ for this example is defined as 
\begin{align}\label{egd}
\rho_c(\lambda) := N|\cos(\lambda-\lambda_{\bf c})| = \begin{cases}
N\cos(\lambda-\lambda_{\bf c}) & \mbox{} \lambda \in {\tilde +}_c \\
-N\cos(\lambda-\lambda_{\bf c}) & \mbox{} \lambda \in {\tilde -}_c
\end{cases}
\end{align}
where $N$ is an arbitrary normalisation constant. The associated nomalised density is therefore
\begin{align}\label{egd2}
\rho'_c(\lambda) := \begin{cases}
{1\over 4}\cos(\lambda-\lambda_{\bf c}) & \mbox{} \lambda \in {\tilde +}_c \\
-{1\over 4}\cos(\lambda-\lambda_{{\bf c}}) & \mbox{} \lambda \in {\tilde -}_c.
\end{cases}
\end{align}
The factor $\cos(\lambda-\lambda_{\bf c})$ gives the component of ${\bf S}$ in the ${\bf c}$-direction whenever ${\bf S}$ belongs to the northern hemisphere ${\tilde +}_c$. Likewise $-\cos(\lambda-\lambda_{\bf c})$ gives the component of ${\bf S}$ in the direction $-{\bf c}$ whenever ${\bf S}$ belongs to the southern hemisphere ${\tilde -}_c$. Thus, the probability densities to obtain the values $+$ and $-$ are proportional to the component of the spin in the positive and negative direction of the SG alignment. Note that the density in (\ref{egd}) is positive semi-definite {\em on the domain of certainty} ${\tilde \Omega}_c$. Using (\ref{p}) and (\ref{egd}) we obtain the single-particle probabilities 
\begin{align}\label{1}
p_c(\pm)=P_c({\tilde \pm}_c|{\tilde \Omega}_c) =P'_c({\tilde \pm}_c)= {1\over 2}.
\end{align}
The average spin measured when the SG alignment is $c$ is
\begin{align}\label{cav}
\langle C \rangle_c = \int_{{\tilde \Omega}_c} \, d\lambda \rho_c(\lambda) {\tilde C}(\lambda)= \sum_{j=\pm} \int_{{\tilde j}_c} d\lambda \, \rho_c(\lambda) {\tilde C}(\lambda) = 0.
\end{align}
The probabilities $p_c(\pm) = 1/2$ are the same as the quantum probabilities
\begin{align}\label{qm}
p_\psi(\pm) := {1\over 2}\left(1\mp {\bf c}\cdot \langle {\bm \sigma}^1 \otimes I^2\rangle_\psi\right) ={1\over 2}
\end{align}
where ${\bm \sigma}^1$ denotes the Pauli operator-valued three-vector for the particle being considered, and $I^2$ denotes the 2-dimensional identity operator on the Hilbert space of a second particle. The average in (\ref{qm}) is taken in the two-particle entangled (Bell) state
\begin{align}\label{bls}
\ket{\psi} := {1\over \sqrt{2}}\left(\ket{\uparrow_{\rm S}}^1\otimes\ket{\downarrow_{\rm S}}^2-\ket{\downarrow_{\rm S}}^1\otimes\ket{\uparrow_{\rm S}}^2\right)
\end{align}
and like $\langle C\rangle_c$ in (\ref{cav}) it is equal to zero.

In this example there is a clear distinction between the spin of the particle in the ${\bf c}$-direction $S_{\bf c}$ and the spin {\em observable} in the ${\bf c}$-direction, which is ${\tilde C}\neq S_{\bf c}$. This is because what is {\em observed} in the experiment depends on the entire experimental arrangement, including the measuring device. The observable $C$ represents a joint system-apparatus observable and the results are ``{\em regarded as the joint product of ‘system’ and ‘apparatus’...}"\cite{bell_speakable_2004}. It is interesting to note that in a sense this is also how the quantum formalism works. The state of the particle pair is given by (\ref{bls}) with the label ${\bf S}$ referring to the particles, but with no explicit reference being made to the alignment $c$ of either SG device. Meanwhile the observables (projections) $\pi_{\pm {\bf c}} := (1\mp {\bf c}\cdot {\bf \sigma})/2$ clearly refer to both a particle {\em and} its measuring device.

Let us now consider the full two-particle EPRB setup. Along with the first SG device measuring the left-particle, we consider a second SG device with possible alignments $b,b'$ measuring the right-particle. The total spin of the particle pair is zero, so the right-particle has spin $-{\bf S}$, which is opposite to that of the left-particle. Thus, for the right-particle the spin observable in the context $c$ has value $+$ when $-{\bf S}$ has positive component in the direction ${\bf c}$, and has value $-$ otherwise;
\begin{align}
{\tilde C}(\lambda)=\begin{cases}
+ & \mbox{} \lambda -\lambda_{-\bf c} \in \left[{-\pi/2}, {\pi/2}\right) \\
- & \mbox{} \lambda -\lambda_{\bf c} \in \left[{-\pi/2}, {\pi/2}\right)
\end{cases}
\end{align}
where $C=B,B'$. The sets ${\tilde \pm}_c,~c=b,b'$ are given by
\begin{align}\label{til2}
{\tilde +}_c = \left[\lambda_{\bf c}+{\pi\over 2},\lambda_{\bf c}+{3\pi\over 2}\right),\qquad
{\tilde -}_c = \left[\lambda_{\bf c}-{\pi\over 2},\lambda_{\bf c}+{\pi\over 2}\right).
\end{align}
We now define the normalised joint-context probability densities $\rho'_{cc'}$ with $c=a,a'$ and $c'=b,b'$ by
\begin{align}\label{2pden}
\rho'_{cc'}(\lambda) = \alpha\rho'_c(\lambda) + \beta \rho'_{c'}(\lambda)
\end{align}
where $\alpha,\beta \in [0,1]$ are constants such that $\alpha+\beta=1$. For a particular choice of $\alpha$ and $\beta$ the above definition uniquely specifies $\rho'_{cc'}$ on the whole domain ${\tilde \Omega}_c\cap{\tilde \Omega}_{c'}$. The fact that $\rho'_{cc'}$ can be expressed as a normalised combination of single-particle densities seems similar to the quantum superposition principle, but actually apart from the requirement that $\alpha+\beta=1$ one has complete freedom in the choice of $\alpha$ and $\beta$.


Despite the fact that the EPRB-setup involves two particles, there is only one unknown quantity $\lambda$. The probability $p_{ab}(j,k)$ can be thought of as the probability that $\lambda$ belongs to both ${\tilde j}_a$ {\em and} ${\tilde k}_b$, so it is obtained by integrating the density $\rho'_a$ over the intersection ${\tilde j}_a\cap{\tilde k}_b$. Since $\rho'_b$ is obtained from $\rho_a'$ by a shift in coordinate axes and an alternative choice of $\alpha$ and $\beta$, $p_{ab}(j,k)$ can also be obtained by integrating $\rho'_b$ over ${\tilde j}_a\cap{\tilde k}_b$. To understand this freedom note that since the context $ab$ is the simultaneous realisation of contexts $a$ and $b$ the observables $A$, $B$ and $AB$ are trivially compatible. We can therefore say that the contexts $a,b$ and $ab$ are themselves compatible---they can be simultaneously realised. Indeed $ab$ can be viewed as a restriction of either one of the contexts $a$ or $b$ due to the other one. This is reflected by the domains of certainty through the relations ${\tilde \Omega}_a\cap{\tilde \Omega}_b \subset {\tilde \Omega}_a$ and ${\tilde \Omega}_a\cap{\tilde \Omega}_b \subset {\tilde \Omega}_b$. The densities $\rho_a'$ and $\rho_b'$ quantify the (Kolmogorovian) epistemic uncertainty in $\lambda$ within the contexts $a$ and $b$ respectively, i.e., over the domains ${\tilde \Omega}_a$ and ${\tilde \Omega}_b$ respectively. Since $\rho_a'$ and $\rho_b'$ quantify the same uncertainty relative to different domains, the joint probabilities found using $\rho_a'$ and $\rho_b'$ should coincide on the common domain ${\tilde \Omega}_a\cap{\tilde \Omega}_b$. Because this is the case $\rho'_{ab}$ is uniquely interpretable as quantifying the uncertainty in $\lambda$ within the context $ab$.

In the case $c=a$, $c'=b$ (\ref{2pden}) yields
\begin{align}\label{v2}
&p_{ab}(+,+)={1\over 4}\left[1-\cos(\lambda_{\bf a}-\lambda_{\bf b})\right]=p_{ab}(-,-)\nonumber \\
&p_{ab}(+,-)={1\over 4}\left[1+\cos(\lambda_{\bf a}-\lambda_{\bf b})\right]=p_{ab}(-,+),
\end{align}
which are the same as the quantum predictions obtained using the Bell state (\ref{bls}). The average $\langle AB\rangle_{ab}$ can be calculated from the probabilities above as
\begin{align}
\langle AB\rangle_{ab} &= \sum p_{ab}(j=k) - \sum p_{ab}(j\neq k)
\nonumber \\ &=-\cos(\lambda_{\bf a}-\lambda_{\bf b}) = -{\bf a}\cdot {\bf b},
\end{align}
which is the same as the corresponding quantum expectation value. It should be clear from this that the present model completely reproduces the quantum predictions, but is also completely classical and deterministic.

We note finally that if one chooses the normalisation $N=1/8$ in (\ref{egd}) then $P_c({\tilde \Omega}_c)=1/2$ can be viewed as the probability that the SG device setting $c$ is selected. Similarly one can define $P_{cc'}$ such that $P_{cc'}({\tilde \Omega}_c\cap{\tilde \Omega}_{c'})=P_c({\tilde \Omega}_c)P_{c'}({\tilde \Omega}_{c'})=1/4$ can be viewed as the probability that $c$ and $c'$ are selected. This normalisation is appropriate for modeling Aspect's experiment in which switches randomly flip between polarisers $a$ and $a'$ as well as between $b$ and $b'$ during the propagation of the light measured \cite{aspect_experimental_1982}. The probabilities in (\ref{1}) and (\ref{v}) can then be understood as the conditional probabilities for outcomes given particular device settings are selected.

\subsection{On the question of locality}

We consider here the status of the above model with regard to the question of locality.

\subsubsection{Bell's locality}

Bell's locality assumption is satisfied in the form 
\begin{align}
p_{ab}(j,k|\lambda) =\delta_{{\tilde j}_a\cap{\tilde k}_b}(\lambda)= \delta_{{\tilde j}_a}(\lambda)\delta_{{\tilde k}_b}(\lambda)=p_a(j|\lambda)p_b(k|\lambda)
\end{align}
where the sets ${\tilde \pm}_a$ are defined in (\ref{til}), and the sets ${\tilde \pm}_b$ are defined in (\ref{til2}). The above is of course just a special case of (\ref{ours}), which holds generally within our contextual framework.

\subsubsection{Einstein locality}

We comment now on another notion of locality. The model above could be deemed to be local if in the two-particle case, all single-particle probabilities are identical to those obtained in the single-particle case. This condition is called {\em Einstein causality}, also known as {\em parameter independence} \cite{fry_experimental_2009}, {\em Einstein locality} \cite{haag_local_1996}, or simply {\em locality} \cite{haag_local_1996,jarrett_separability_2009}. It refers to the impossibility of superluminal signaling. Parameter independence is satisfied within the above model. To see this note that the single-particle probabilities in (\ref{1}) refer to a particle for which the spin is measured along the $a,a'$ directions, with no reference being made to any other particles or any other SG devices. In the model of the full two-particle EPRB-setup, single-particle probabilities for the left-particle are given for $c=a,a'$ by
\begin{align}\label{v}
&p_c(j|\lambda) = \sum_{k=\pm} p_{cb}(j,k|\lambda) = \sum_{k=\pm} p_{cb'}(j,k|\lambda),\nonumber \\
&p_c(j) = \sum_{k=\pm} p_{cb}(j,k) = \sum_{k=\pm} p_{cb'}(j,k),
\end{align}
and the probabilities $p_c(j)$ above are identical to those found in (\ref{1}). Thus, the probabilities pertaining to the left-particle obtained by considering the full two-particle setup, are identical to those obtained in the case that no right-particle exists. In particular the probabilities on the left-hand-side in (\ref{v}) are independent of the context label $c'=b,b'$ chosen for the right-particle. The above result is analogous to the ``no-signalling" theorems of quantum theory \cite{peres_quantum_2004}. 

\subsubsection{Discussion}

There are numerous situations in classical physics in which spacelike separated events are statistically correlated. This instantaneous action-at-a-distance is associated with logical inference rather than causal influence, so no violations of causality ensue. For example, if the spin of the left-particle ${\bf S}$ is found to be $+$ in the direction $a$, then we know with certainty that the spin of the right-particle is $-$ in the direction $a$, and this logical inference propagates instantaneously. {\em Outcome independence} is the condition of statistical independence given the underlying ontic state $\lambda$;
\begin{align}
p^1_\lambda(j|a)=p_\lambda(j|a,k),\qquad p^2_\lambda(k|b) = p_\lambda(k|j,b)
\end{align}
where we have again borrowed the notation from \cite{fry_experimental_2009}. Together, parameter independence and outcome independence are equivalent to Bell's locality assumption \cite{jarrett_separability_2009}.

Any theory consistent with special relativity must satisfy Einstein causality. However, it appears that quantum theory violates outcome independence, where classical theory cannot. This means that quantum theory is nonlocal in a sense that classical theory is not. In the Copenhagen interpretation of quantum theory, a particle can only be said to possess a definite spin value if it is in an eigenstate of the appropriate spin operator. In this sense quantum theory is {\em indeterministic}. As a result, in quantum theory a violation of outcome independence occurs when a measurement is made on the left-particle in the Bell state (\ref{bls}), because this measurement {\em physically determines} the spin of the right-particle through the collapse of the composite superposition (\ref{bls}). Prior to the measurement of the left-particle, the right-particle could not be said to possess a definite spin value, because the Bell state (\ref{bls}) is not an eigenstate of the right-particle's spin operator. This nonlocal action is viewed as the cause of the correlations predicted, which cannot then be interpreted as due to mere logical inferences. This is what Einstein described as ``spooky action-at-a-distance".

Since experiments violate the CHSH inequalities, it is alleged that the ``spooky action-at-a-distance" of quantum theory {\em must} be accepted. There are two fallacies from which this conclusion results. The first is that Bell-type inequalities rest upon Bell's locality assumption, and in particular on outcome independence. The second is that classical theories cannot violate Bell-type inequalities. Of course, quantum theory is (by definition) non-classical. What is more, in the Copenhagen interpretation it is not deterministic, and it violates outcome independence. It also happens to violate Bell-type inequalities. However, this does {\em not} imply that the {\em only} way to violate Bell-type inequalities is to use a non-classical, nonlocal, indeterministic theory. Actually as we have shown, one can reproduce the quantum predictions with a completely deterministic, local, classical theory, provided one adopts a sufficiently refined view towards measurement. As a result the fallacious conclusion that Bell's inequalities necessarily impose constraints on classical physics should be summarily disregarded.

The example above is purely classically deterministic, so one can argue that in this context the only meaningful notion of locality is Einstein causality, i.e., parameter independence. One does not need to distinguish between two types of locality, and explanations for certain correlations as resulting from a ``spooky action-at-a-distance" are unnecessary. Having eliminated the false idea that classical correlations cannot violate Bell-type inequalities one is free to interpret the correlations observed in a purely classical way.

\section{Conclusions}\label{conc}

We have introduced an alternative probabilistic model of EPRB-type experiments using multi-valued maps of the kind found in Dempster-Shafer probability theory \cite{shafer_mathematical_1976,yager_classic_2008}. This has lead to a contextual approach, which appears to be similar in nature to Khrennikov's \cite{khrennikov_contextual_2009,khrennikov_classical_2014}. Our approach gives an explicit link between measurement contexts and counterfactual outcomes. The uncertainty associated with the latter is non-Kolmogorovian epistemic uncertainty in the underlying event space $\Omega$, which is divided up according to different measurement contexts using the multi-valued maps. These concepts have been applied to both phenomenological and hidden-variable treatments of Bell-type inequalities. In both cases the counterfactual assumption is seen to be invalid. The ensuing non-Kolmogorovian epistemic uncertainty leads to context conditioned probabilities that do not obey Bell-type inequalities.

The conditioning of probabilities on measurement contexts can be viewed as the result of a formulation in which the measurement devices are not {\em passive}, but {\em active} in determining physical events. This idea of what should be taken to constitute a ``measurement" was advocated by Bell long ago \cite{bell_speakable_2004}. What appears not to have been recognised by many physicists is that this has implications for classical as well as quantum physics. It implies that the restriction to a single Kolmogorov probability space should {\em not} be viewed as a {\em necessity} for a classical description. Kolmogorov probability theory and classical physics are two distinct theories---one is a theory of probability and one is a theory of physics. Although in some cases they may be inter-related, it is a mistake to conflate assumptions about the former with assumptions about the latter. This is precisely the mistake one makes in concluding that Bell-type inequalities constrain classical physics.

While the context-conditioned probabilities we have obtained do not obey Bell-type inequalities, there is nothing about our approach which precludes classical, deterministic, local physical models. In particular we have constructed a local and deterministic classical model, which reproduces the quantum-mechanical predictions. From this point of view, the incorrect conclusion that violations of Bell-type inequalities rule out deterministic, local, classical theories simply results from an oversimplified conception of measurement.

{\em Acknowledgment}. I would like to thank T. Barlow, R. Bennett, P. Knott, and T. Proctor for useful discussions relating to this work.

\bibliography{b.bib}

\end{document}